\documentstyle[12pt,epsf]{article}

\def\ffrac#1#2{\textstyle{#1\over#2}\displaystyle}

\begin{document}
\title{The O($n$) loop model on the 3-12 lattice} 
\author{M. T. Batchelor\\
{\small
Department of Mathematics, School of Mathematical Sciences},\\
{\small 
The Australian National University, Canberra ACT 0200, Australia}}
\date{May 21, 1998}
\maketitle

\begin{abstract}
The partition function of the O($n$) loop model on the honeycomb
lattice is mapped to that of the O($n$) loop model on the 
3-12 lattice. Both models share the same operator 
content and thus critical exponents. The critical points are 
related via a simple transformation of variables. 
When $n=0$ this gives the recently found exact value 
$\mu = 1.711~041\ldots$ for the connective constant of 
self-avoiding walks on the 3-12 lattice.
The exact critical points are recovered for the Ising model on 
the 3-12 lattice and the dual asanoha lattice at $n=1$.   
\end{abstract}

\section{Introduction}

Consider the O($n$) loop model on the honeycomb lattice with 
${\cal N}_h$ vertices. The partition function is defined by 
\begin{equation}
{\cal Z}_h = \sum_{\rm loops} t^{{\cal N}_h - L} n^P,
\end{equation}
where the sum is over all configurations of non-intersecting loops.
In any given configuration $L$ is the number of occupied vertices, with
$t^{{\cal N}_h - L}$ the weight of the empty (unoccupied) vertices. 
The variable $n$ is the fugacity of a closed loop, with $P$ their 
number in any given configuration. A typical loop configuration is shown 
in Fig. 1. 

\begin{figure}
\centerline{
\epsfxsize=2.5in
\epsfbox{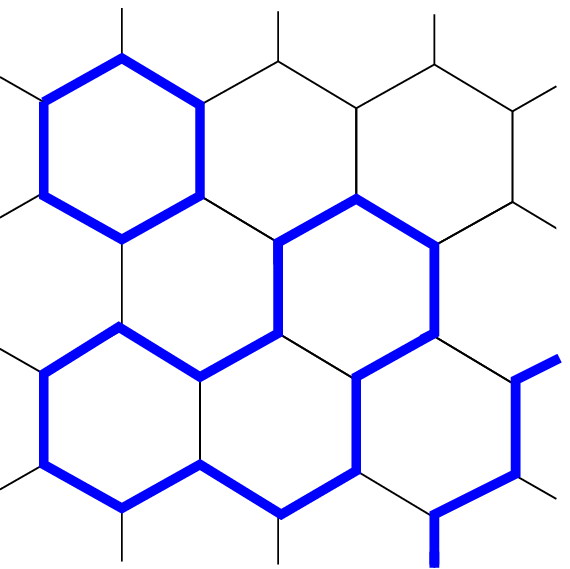}}
\caption{Loop configuration on the honeycomb lattice with weight $t^5 n^2$.}
\end{figure}

This model originates from the high-temperature expansion \cite{DMNS} 
of the O($n$) or $n$-vector model \cite{S}. Nienhuis identified
two branches of critical points for the O($n$) loop model on the 
honeycomb lattice defined by \cite{N}
\begin{equation}
(2-t^2)^2  = 2 - n \, .   \label{poly}
\end{equation}
Baxter has shown that this is also the necessary condition for which 
the row transfer matrix of the corresponding 3-state vertex model can 
be diagonalised by means of the co-ordinate Bethe Ansatz \cite{B86}. 
This vertex model turned out to be a special case of a more general 
solvable model defined on the square lattice \cite{IK}. 
The Bethe Ansatz solution was later extended 
to open boundary conditions \cite{BS,YB}.

Most importantly, the O($n$) loop model on the honeycomb lattice
has led to a wealth of exact information on the configurational 
properties of self-avoiding walks in the $n\to 0$ \cite{dG} limit. 
For a review, we refer the reader to \cite{D}. The simplest result
concerns the enumeration of a single $N$-step self-avoiding walk from a 
point deep in the bulk, for which the number of configurations
scales as  
\begin{equation}
C_N \sim \mu^N N^{\gamma-1}
\end{equation}
for large $N$. The connective constant, 
$\mu = \sqrt{2+\sqrt{2}} = 1.847~759\ldots$, follows from (\ref{poly}).
The configurational exponent $\gamma=\frac{43}{32}$ was first
obtained via Coulomb gas calculations \cite{N}.   
More general configurational exponents have been obtained for
arbitrary networks of long chains, both in the bulk and near a
surface \cite{D}.\footnote{The most recent developments are reported
in \cite{BBO}.}  

Given the solvability of the O($n$) loop model on the honeycomb lattice
along the line of critical points defined by (\ref{poly}),
I had often wondered if the corresponding model 
could be solved on the 3-12 lattice depicted in Fig. 2.
It also has co-ordination number three.
Prompted by a suggestion that 
the connective constant for self-avoiding walks on the
3-12 lattice may also be obtained exactly,
Jensen and Guttmann have found the value $\mu = 1.711~041\ldots$ \cite{JG}.
They were able to relate the generating functions for both
self-avoiding walks and self-avoiding polygons on the
honeycomb lattice to those on the 3-12 lattice
by a simple change of variables. This mapping is discussed
here in the context of the more general O($n$) loop model,
which indeed turns out to be solvable at criticality. 

\begin{figure}
\centerline{
\epsfxsize=2.5in
\epsfbox{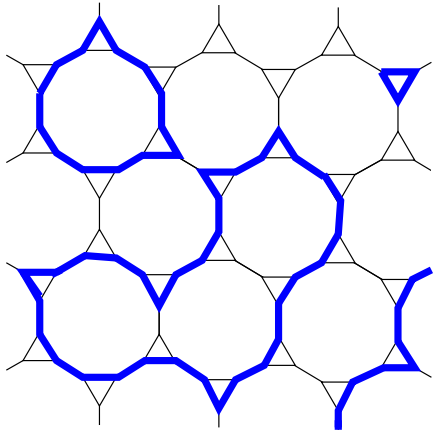}}
\caption{Loop configuration on the 3-12 lattice.}
\end{figure}

\section{O($n$) loop model on the 3-12 lattice}

Vertices on the honeycomb lattice are either empty or occupied.
For each type of vertex configuration on the honeycomb lattice
there are two possible configurations on the 3-12 lattice,
as shown in Fig. 3. It thus follows that any given configuration
of loops on the honeycomb lattice, with weight $t^{{\cal N}_h - L} n^P$,
maps to $(t^3 + n_\Delta)^{{\cal N}_h - L} (t+1)^L n^P$ possible 
configurations on the 3-12 lattice, with $n_\Delta=n$. 
One of the $(t^3 + n_\Delta)^5 (t+1)^{23} n^2$ possible
configurations generated from the honeycomb configuration in
Fig. 1 is shown in Fig. 2. 

\begin{figure}
\centerline{
\epsfxsize=3.5in
\epsfbox{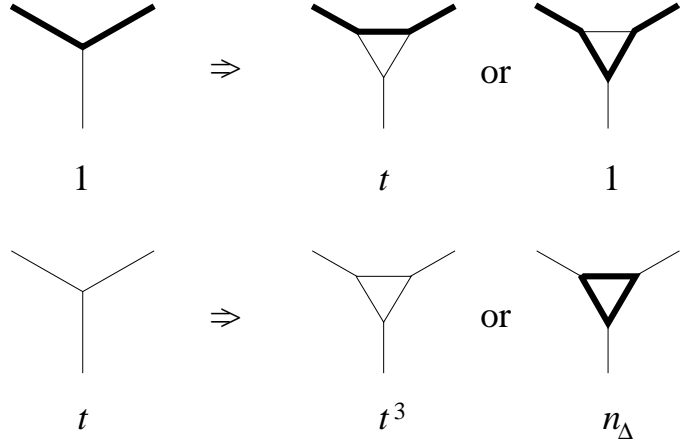}}
\caption{Mapping between vertex configurations. The other possible vertex
weights are similarly defined under uniform rotation.}
\end{figure}

The partition function of the O($n$) loop model on the 3-12 
lattice can be written 
\begin{eqnarray}
{\cal Z}_{3-12} &=& \sum_{\rm loops} 
(t^3 + n_\Delta)^{{\cal N}_h - L} (t+1)^L n^P \nonumber\\
&=& (t^3 + n_\Delta)^{{\cal N}_h} \sum_{\rm loops} 
\left(\frac{t+1}{t^3 + n_\Delta}\right)^L n^P \, .
\end{eqnarray}
This is to be compared with
\begin{equation}
{\cal Z}_h = t^{{\cal N}_h} \sum_{\rm loops} 
\left(\frac{1}{t}\right)^L n^P \, .
\end{equation}

It follows that the critical points of the O($n$) loop model on the 
3-12 lattice are given by solving (\ref{poly}) with
\begin{equation}
\frac{1}{t} \to  \frac{t+1}{t^3 + n} \, .
\label{map}
\end{equation}

\subsection{Self-avoiding walks}

In particular, when $n=0$
\begin{equation}
  2 + 8\,t + 12\,{t^2} + 8\,{t^3} + 2\,{t^4} - 4\,{t^6} - 8\,{t^7} - 
   4\,{t^8} + {t^{12}} = 0  \, .
\end{equation}
The exact connective constant $\mu=1.711~041\ldots$ follows from
the largest real root, as obtained by Jensen and Guttmann \cite{JG}.
Equivalently, it follows on solving 
\begin{equation}
\frac{1}{\sqrt{2+\sqrt 2}} = \frac{1}{t^2} + \frac{1}{t^3}.
\end{equation}
Their mapping $x \to x^2 + x^3$ between generating functions
follows on defining $x=1/t$ as the weight per step.   

\subsection{The Ising model}

Another point of note is the Ising model at $n=1$. It is known that
the critical
temperature of the Ising model on both the honeycomb and triangular 
lattices follows from (\ref{poly}) and the standard duality relation.
In a similar way, the critical Ising point can be obtained from (\ref{map})
for both the 3-12 lattice and its dual (the asanoha lattice).    
For $n=1$ (\ref{poly}) and (\ref{map}) give
\begin{equation}
\frac{1}{\sqrt 3} = \frac{1}{t^2-t+1}.
\end{equation}
Taking the positive root gives
\begin{equation}
{\rm e}^{2 A_c} = \ffrac{1}{2}\left( 1 + \sqrt{4 \sqrt{3} - 3} \,\right)
= 1.490~984\ldots
\label{asa}
\end{equation}
as the critical coupling on the dual asanoha lattice. The critical
point
\begin{equation}
{\rm e}^{2 K_c} = \ffrac{1}{2}\left( 3 + \sqrt{3} + 
\sqrt{12 + 10 \sqrt{3}} \,\right) = 5.073~446\ldots
\label{3.12}
\end{equation}
on the 3-12 lattice follows from the duality relation
${\rm e}^{-2 A_c} = \tanh K_c$.
The Ising values (\ref{asa}) and (\ref{3.12}) are precisely 
those given by Syozi \cite{Sy}, who arrived at the
Ising model on the 3-12 lattice from the
Ising model on the honeycomb lattice after the successive application
of the double decoration process and the star-triangle transformation.

\section{Conclusion}

The partition function of the O($n$) loop model on the honeycomb
lattice has been mapped to that of the O($n$) loop model on the 
3-12 lattice. The critical behaviour of both models is thus 
related. In particular, they share the same operator content and
thus critical exponents. 
Although the mapping between the models is particularly simple,
it nevertheless provides a clear example of universality
between models defined on regular and semi-regular lattices.
The non-universal features, such as the
critical points, are related via the transformation (\ref{map}). 
When $n=0$ this gives the exact value $\mu = 1.711~041\ldots$
found recently by Jensen and Guttmann \cite{JG} for the
connective constant of self-avoiding walks on the 3-12 
lattice.\footnote{Other non-universal quantities for
self-avoiding walks on the 3-12 lattice 
can also be obtained from their honeycomb counterparts, such as
the critical adsorption temperature at a boundary \cite{BY}.}
When $n=1$ the exact critical points, (\ref{3.12}) and (\ref{asa}),
are recovered for the Ising model on the 3-12 lattice 
and the dual asanoha lattice.

\section{Acknowledgements}
It is a pleasure to thank Tony Guttmann and Ivan Jensen for their
interest in the self-avoiding walk problem and for communicating 
their results prior to publication. I gratefully acknowledge financial 
support from the Australian Research Council.

\end{document}